\documentclass[a4paper,11pt]{amsart}
\usepackage{amsfonts,amssymb,epsfig,latexsym}
\usepackage{amsmath}
\usepackage[latin1]{inputenc}
\usepackage{color,layout}
\usepackage{ifthen}

\usepackage{graphicx}
\usepackage{color}

\newtheorem{theorem}{Theorem}[section]

\newtheorem{proposition}[theorem]{Proposition}

\def\square{\hbox{\vrule\vbox{\hrule\phantom{o}\hrule}\vrule}}

\def\eps{\varepsilon}

\newcommand{\be}{\begin{equation}}
\newcommand{\ee}{\end{equation}}

\newcommand{\texr}{\textcolor{red}}

\numberwithin{equation}{section}

\newcommand{\Z}{\mathbb{Z}}
\newcommand{\R}{\mathbb{R}}
\newcommand{\C}{\mathbb{C}}
\newcommand{\W}{{\mathcal W}}

\newcommand{\p}{\partial}

\newcommand{\lan}{\langle}
\newcommand{\ran}{\rangle}
\newcommand{\pal}{\parallel}

\newcommand{\e}{\varepsilon}

\newcommand{\re}{{\rm Re}}
\newcommand{\im}{{\rm Im}}
\newcommand{\ord}{{\mathcal O}}
\newcommand{\ai}{{\rm Ai}\,}
\newcommand{\bi}{{\rm Bi}\,}

\def\eq#1{(\ref{#1})}

\newtheorem{thm}{Theorem}[section]

\numberwithin{equation}{section}

\begin{document}

\title[Resonances near an energy-level crossing II]
{Molecular predissociation  resonances near an energy-level crossing II: \\
Vector field interaction}
\author{S.~Fujii\'e${}^1$, A.~Martinez${}^2$ and T.~Watanabe${}^3$}



\maketitle
\addtocounter{footnote}{1}
\footnotetext{{\tt\small  Department of Mathematical Sciences, Ritsumeikan University,  
fujiie@fc.ritsumei.ac.jp} }  
\addtocounter{footnote}{1}
\footnotetext{{\tt\small Universit\`a di Bologna,  
Dipartimento di Matematica, 
andre.martinez@unibo.it }}  
\footnotetext{{\tt\small  Department of Mathematical Sciences, Ritsumeikan University,  
t-watana@se.ritsumei.ac.jp} }

\begin{abstract}
We study the resonances of a two-by-two semiclassical system of one dimensional Schr\"odinger operators, 
near an energy where the two potentials intersect transversally, one of them being bonding, 
and the other one anti-bonding. Assuming that the interaction is a vector-field, 
we obtain optimal estimates on the location and on the widths of these resonances.
\end{abstract}
\vskip 2cm


\section{Introduction}
This paper is devoted to the study of diatomic molecular predissociation resonances in the Born-Oppenheimer approximation, at energies close to that of the crossing of the electronic levels.  It is a continuation of \cite{FMW} where a method was introduced in order to overcome the difficulty of working with a $2\times 2$ system of semiclassical operators. 

In all of the work, the parameter $h$ stands for the square-root of the inverse of the (mean-) mass of the nuclei. 

In \cite{FMW}, we obtained  optimal estimates both on the real parts and on the imaginary parts (widths) of the resonances (that respectively correspond to the radiation frequency and to the inverse of the life-time of the molecule), under the condition that the interaction is of the form $h(r_0(x)+hr_1(x)D_x)$, with $r_0\not=0$ at the point where the two electronic levels cross. However, when performing a Fechbach reduction in the Born-Oppenheimer approximation (see, e.g., \cite{KMSW, MaMe, MaSo}), it appears that the interaction that comes out is a vector-field of the form $ih^2r_1(x)D_x$ (plus smaller terms), with $r_1$ real on the real. In that case, the result of \cite{FMW} just says that the widths are $\ord (h^2)$, and does not provide any lower bound on them.

Here we plan to apply the techniques introduced in \cite{FMW} in order to obtain the asymptotic behaviour of the widths of the resonances, in the physical case of a vector-field interaction.

As in \cite{FMW}, we consider a $2\times2$ matrix system, the diagonal part of which consists of one-dimensional semiclassical Schr\"odinger operators, and we assume that the two potentials cross transversally at the origin, with value 0, and that, at this energy level, one of the two potentials admits a well, while the other one is non-trapping (see figure 1).

\begin{figure}[h]
\label{fig1}
\begin{center}
\scalebox{0.5}[0.35]{
\includegraphics{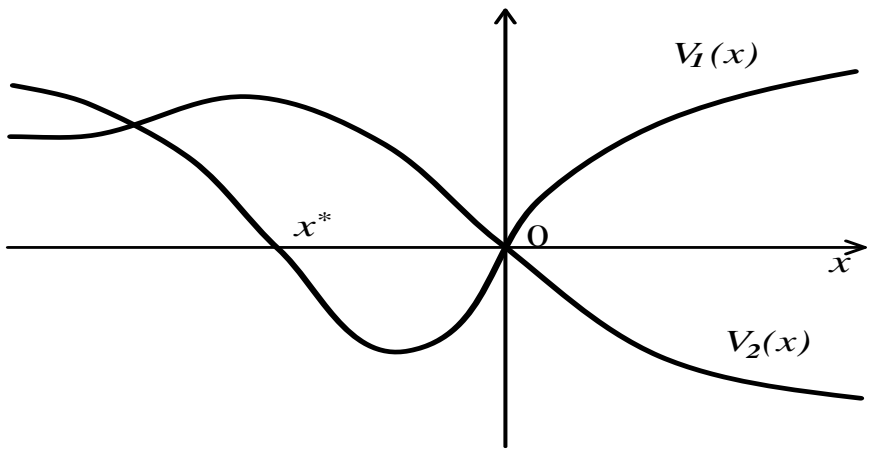}
}
\end{center}
\caption{The two potentials}
\end{figure}

For such a model, we study the resonances $E=E(h)$ that have a real part $\ord (h^{2/3})$ and an imaginary part $\ord (h)$.

\section{Assumptions and results}

We consider a Schr\"odinger operator with  $2\times 2$ matrix-valued potential,
\begin{equation}
\label{sch}
\begin{aligned}
Pu &= Eu,\qquad
P &= \left(
\begin{matrix}
 P_1 & hW\\
hW^* & P_2
\end{matrix}
\right),
\end{aligned}
\end{equation}
where $P_j=h^2D_x^2 + V_j(x)$ ($j=1,2$) with $D_x=-i\frac{d}{d x}$, , $W=W(x,hD_x)$ is a first order semiclassical differential operator, 
and $W^*$ is the formal adjoint of $W$.

As in \cite{FMW}, we suppose the following conditions on the potentials $V_1(x), 
V_2(x)$ and on the interaction
$W(x,hD_x)$:

{\bf (A1)}
$V_1(x)$, $V_2(x)$ 
 are real-valued analytic functions on $\R$, and extend to holomorphic functions
 in the complex domain,
$$
\Gamma=\{x\in\C\,;\,|\im\, x|<\delta_0\lan\re \,x\ran\}
$$
where $\delta_0>0$ is a constant, and $\lan t\ran:=(1+|t|^2)^{1/2}$.

{\bf (A2)} For $j=1,2$, $V_j$ admit limits as $\re\, x\to \pm\infty$ in $\Gamma$, and they satisfy,
$$
\begin{aligned}
\lim_{{\re\,x\to -\infty}\atop{x\in \Gamma}}  V_1(x)>0\, ;\, \lim_{{\re\,x\to -\infty}\atop{x\in \Gamma}} V_2(x)>0\, ;\\
\lim_{{\re\,x\to +\infty}\atop{x\in \Gamma}} V_1(x)>0\, ;\, \lim_{{\re\,x\to +\infty}\atop{x\in \Gamma}} V_2(x)<0.
\end{aligned}
$$

{\bf (A3)} There exists a negative number $x^*<0$ such that,
\begin{itemize}
\item $V_1>0$ and $V_2>0$ on $(-\infty, x^*)$;
\item $V_1<0<V_2$ on $(x^*,0)$;
\item $V_2<0<V_1$ on $(0,+\infty)$,
\end{itemize}
and one has,
$$
V_1'(x^*)  <0,\quad  V_1'(0)=:\tau_1>0,\qquad V_2'(0)=:-\tau_2<0.
$$

{\bf  (A4)}
The intraction $W(x,hD_x)$ is a differential operator of the form,
$$
W(x,hD_x) = r_0(x) + ir_1(x)hD_x,
$$
where $r_0(x)$ and $r_1(x)$ are bounded analytic functions on $\Gamma$, 
and $r_0(x)$ is real-valued on $\R$.

\vskip 0.3cm

Notice that, in a neighborhood of $E=0$, the scalar operator $P_1$ has eigenvalues, 
while $P_2$ has only essential spectrum. Hence, if the interaction $W$ is absent, 
the matrix-valued operator $P$ 
has embedded eigenvalues in the essential spectrum. But if $W$ is present, 
it is expected that there exist, instead of embedded eigenvalues,  resonances close to them in the lower
half complex plane of the energy.

The resonances of $P$ are defined, 
e.g., as the values $E\in\C$ such that the equation $Pu=Eu$ has a non identically vanishing solution such that, 
for some $\theta >0$ sufficiently small, the function $x\mapsto u(xe^{i\theta})$ 
is in $L^2(\R)\oplus L^2(\R)$ (see, e.g., \cite{AgCo, ReSi}). 
We will give an equivalent definition of resonances adapted to our setting in the next section
 (see also \cite{HeMa}). 
We denote by ${\rm Res}(P)$ the set of these resonances.

For $E\in \C$ small enough, we define the action,
\begin{equation}
\label{action}
{\mathcal A}(E):= \int_{x_1^*(E)}^{x_1(E)}\sqrt{ E-V_1(t)} \, dt,
\end{equation}
where $x_1^*(E)$ (respectively $x_1(E)$) is the unique solution of $V_1(x)=E$ close to $x^*$ (respectively close to 0).

We fix $C_0>0$ arbitrarily large, and we study the resonances of $P$ lying in the set ${\mathcal D}_h(C_0)$ given by,
\be
{\mathcal D}_h(C_0):= [-C_0h^{2/3}, C_0h^{2/3}]-i[0,C_0h].
\ee
For $h>0$ and $k\in\Z$, we set,
\be
\label{deflambdakh}
\lambda_k(h):=\frac{-{\mathcal A}(0)+(k+\frac 12)\pi h}{{\mathcal A}'(0)h^{2/3}}.
\ee
The Bohr-Sommerfeld quantization condition of eigenvalues for the scalar operator $P_1$
reads
$$
{\mathcal A}(E)=(k+\frac 12)\pi h+\ord (h^2).
$$
Then the $\lambda_k(h)h^{2/3}$'s are approximate eigenvalues of $P_1$ near 0. We will find resonances close to these 
real values.

First we recall the result obtained in \cite{FMW}. 
Notice that a multiplicative factor
$(\tau_1\tau_2)^{\frac{1}{3}}$  
in the asymptotic formula of the imaginary part of resonances 
was missing in that paper. We would like to correct it on this occasion.

\begin{thm}[\cite{FMW}]\sl
\label{mainth1}
Assume (A1)-(A4). For $h>0$ small enough, one has,
$$
{\rm Res}\,(P)\cap {\mathcal D}_h(C_0)
=\{E_k(h); k\in\Z\}\cap{\mathcal D}_h(C_0),
$$
where the
 $E_k(h)$'s are complex numbers that satisfy,
\be
\label{reEk0}
\re \,E_k(h) = \lambda_k(h) h^{\frac{2}{3}} - \frac{{\mathcal A}''(0)}{2{\mathcal A}'(0)} \lambda_k(h)^2 h^{\frac{4}{3}}
 + \ord (h^{\frac{5}{3}}),
\ee
\be
\label{imEk0}
\im \,E_k(h)= -\frac{2\pi^2 r_0(0)^2}{{\mathcal A}'(0)} (\tau_1\tau_2)^{\frac{1}{3}} \Bigl(\mu_1(\lambda_k(h))^2 + \mu_2(\lambda_k(h))^2 \Bigr) h^{\frac{5}{3}} + \ord(h^2),
\ee
uniformly as $h\to 0_+$. Here, the functions $\mu_1$ and $\mu_2$ are defined by 
\begin{align*}
\mu_1(t) &= \int_0^\infty \ai(\tau_1^{-\frac{2}{3}} (\tau_1 y - t)) \ai(-\tau_2^{-\frac{2}{3}} (\tau_2 y + t))\,dy\\
\mu_2(t) &= \int_0^\infty \ai(\tau_2^{-\frac{2}{3}} (\tau_2 y - t)) \ai(-\tau_1^{-\frac{2}{3}} (\tau_1 y + t))\,dy,
\end{align*}
where $\ai$ stands for the Airy function 
$\ai(x)=\frac 1{2\pi}\int_{-\infty}^\infty e^{i(x\xi+\xi^3/3)}d\xi.$
\end{thm}

Remark that, if $r_0(0)=0$, this theorem
gives only an estimate $\ord (h^2)$ for the imaginary part of resonances.
The following main result provides a precise asymptotic formula up to $\ord (h^{\frac 83})$
in  the case where $r_0(x)$ vanishes
identically.

\begin{thm}\sl
\label{mainth2}
Assume moreover that $r_0(x)=0$ and $r_1(x)$ is real on $\R$. 
Then the
 $E_k(h)$'s in the previous theorem satisfy, with $\tau_3^{-1}=\tau_1^{-1} + \tau_2^{-1}$
\be
\label{reEk}
\re \,E_k(h) = \lambda_k(h) h^{\frac{2}{3}} - \frac{{\mathcal A}''(0)}{2{\mathcal A}'(0)} \lambda_k(h)^2 h^{\frac{4}{3}}
- \frac{{\mathcal A}^{(3)}(0)}{6{\mathcal A}'(0)}\lambda_k(h)^3h^{\frac{6}{3}} + \ord (h^{\frac{7}{3}}),
\ee
\be
\label{imEk}
\im \,E_k(h)= -\frac{\pi^2 r_1(0)^2}{{\mathcal A}'(0)}\frac{\tau_3^{\frac{1}{3}}}{\tau_1 + \tau_2} 
\left(\ai' (- \tau_3^{-\frac{2}{3}}\lambda_k(h) ) \right)^2 h^{\frac{7}{3}} + \ord(h^{\frac{8}{3}}),
\ee
uniformly as $h\to 0_+$. \end{thm}


\section{Proof of Theorem \ref{mainth2}}
\label{sect3}

We prove Theorem \ref{mainth2}. 
For a sufficiently small $\theta>0$, let $
I_L:= (-\infty, 0]$, $I_R^\theta:=F_\theta ([0, +\infty))$ with
$F_\theta (x):= x+i\theta f(x)$
where $f\in C^{\infty}(\R_+,\R_+)$, $f(x)=x$ for $x$ large enough, $f(x)=0$ for $x\in[0,x_\infty]$ for some $x_\infty>0$, and $f$ is chosen in such a way that, for any $x\geq x_\infty$, and with some positive constant $C$, one has (see \cite{FMW}, Formula (3.1)),
\be
\label{decItheta}
\im \int_{x_\infty}^{F_\theta (x)}\sqrt{ E-V_2(t)} dt \geq -Ch.
\ee

The linear space $V$ of solutions to the system \eq{sch} is of dimension four.
The  solutions in 
$L^2(I_R^\theta)\oplus L^2(I_R^\theta)$ form a two dimensional subspace 
$V_R=V\cap (L^2(I_R^\theta)\oplus L^2(I_R^\theta))$, and 
the solutions in 
$L^2(I_L)\oplus L^2(I_L)$ form a two dimensional subspace 
$V_L=V\cap (L^2(I_L)\oplus L^2(I_L))$.

Then $E$ is a resonance if and only if the intersection $V_R\cap V_L$ is at least 1 dimensional.
In other words, the quantization condition of resonances can be written in the form
\be
\label{wronskian0}
{\mathcal W}_0(E):=
{\mathcal W}(w_{1,L}, w_{2,L}, w_{1,R}, w_{2,R})=0,
\ee
where the couple $(w_{1,L}, w_{2,L})$ (resp. $(w_{1,R}, w_{2,R})$) is a basis of $V_L$ 
(resp. $V_R$)  
and ${\mathcal W}(w_{1,L}, w_{2,L}, w_{1,R}, w_{2,R})$ 
is the Wronskian, i.e. the determinant of the $4\times 4$ matrix
$$
\left (
\begin{array}{cccc}
w_{1,L} & w_{2,L} & w_{1,R} & w_{2,R} \\
\partial_xw_{1,L} & \partial_xw_{2,L} & \partial_xw_{1,R} & \partial_xw_{2,R}
\end{array}
\right ).
$$

Such solutions $w_{1,L}, w_{2,L}, w_{1,R}, w_{2,R}$ are constructed as in \cite{FMW}
using fundamental solutions to the scalar equations $(P_j - E)u = 0$.

On $I_L $, and for $E\in {\mathcal D}_h(C_0)$ and  $j=1,2$, 
let $u_{j,L}^\pm$ be the solutions to 
$(P_j - E)u = 0$ constructed in \cite{FMW} (in particular, $u_{j,L}^-$ 
decays exponentially at $-\infty$, while $u_{j,L}^+$ grows exponentially, and their Wronskian $\W[u_{j,L}^+,u_{j,L}^-]$ is of size $h^{-\frac23}$).
We construct fundamental solutions  $K_{j,L}, \,j=1,2$ on
$I_L$:
\be
\label{eq1}
\begin{aligned}
K_{j,L}[v](x) := \frac{u_{j,L}^+(x)}{h^2 \W[u_{j,L}^+,u_{j,L}^-]} \int_{-\infty}^x & u_{j,L}^-(t)v(t)\,dt \\
& + \frac{u_{j,L}^-(x)}{h^2 \W[u_{j,L}^+,u_{j,L}^-]} \int_x^{0}\!\!\!\! u_{j,L}^+(t)v(t)\,dt.
\end{aligned}
\ee

In the same way, we construct $K_{j,R}, \,j=1,2$ on $I_R^\theta$
using the solutions $u_{j,R}^\pm$ ($u_{j,R}^+$ grows and $u_{j,R}^-$ decays exponentially,
at $\infty$ along $I_R^\theta$):
\be\label{eq1R}
\begin{aligned}
K_{j,R}[v](x) :=  \frac{u_{j,R}^-(x)}{h^2 \W[u_{j,R}^-,u_{j,R}^+]} &\int_{0}^x u_{j,R}^+(t)v(t)\,dt \\
&+ \frac{u_{j,R}^+(x)}{h^2 \W[u_{j,R}^-,u_{j,R}^+]} \int_x^{+\infty}\!\!\!\! u_{j,R}^-(t)v(t)\,dt.
\end{aligned}
\ee

Let
$C^0_b(I_L)$ and $C^0_b(I_R^\theta)$ be the space of bounded continuous functions on $I_L$ and $I_R^\theta$
respectively. The above operators act on these function spaces,  and satisfy 
$(P_j-E)K_{j,L}={\rm Id}$ and $(P_j-E)K_{j,R}={\rm Id}$ respectively.

We have the following estimates, which are better than the elliptic case (see Proposition 3.1 and 3.2 in \cite{FMW}). 

\begin{proposition}\sl
\label{claim} As $h$ goes to $0_+$,
one has uniformly,
\begin{align}\label{est3.1.1}
\pal hK_{2,L}W^* \pal_{{\mathcal L}(C^0_b(I_L))} &= \ord (h^{\frac23}),\\ \label{est3.1.2}
\pal h^2K_{1,L}WK_{2,L}W^* \pal_{{\mathcal L}(C^0_b(I_L))} &= \ord (h),\\ \label{est3.1.3}
\pal hK_{1,R}W \pal_{{\mathcal L}(C_b^0(I_R^\theta))} &= \ord (h^{\frac23}),\\ \label{est3.1.4}
\pal h^2K_{2,R}W^*K_{1,R}W \pal_{{\mathcal L}(C_b^0(I_R^\theta))} &= \ord (h).
\end{align}
\end{proposition}

\noindent{\bf Proof:}\quad 
We first prove the estimates for $K_{j,L}$.
For $j=1,2$, we set,
\be
\begin{aligned}
\label{defUj}
& U_j(x,t):= |u_{j,L}^+(x)u_{j,L}^-(t)|{\bf 1}_{\{t<x\}}+  |u_{j,L}^-(x)u_{j,L}^+(t)|{\bf 1}_{\{t>x\}}=U_j(t,x);\\
&U_j'(x,t):= |u_{j,L}^+(x)\partial_tu_{j,L}^-(t)|{\bf 1}_{\{t<x\}}+  |u_{j,L}^-(x)\partial_tu_{j,L}^+(t)|{\bf 1}_{\{t>x\}};\\
& \widetilde U_j(x,t):=U_j(x,t)+U_j'(x,t).
\end{aligned}
\ee
Thanks to our choice of $K_{j,L}$, and doing an integration by parts,  we see that,
\be
\label{estK2L}
\begin{aligned}
& |hK_{1,L}Wv(x)|=\ord (h^{\frac23})\left( \int_{-\infty}^0\widetilde U_1(x,t)|v(t)|dt + U_1(x,0)|v(0)|\right);\\
& |hK_{2,L}W^*v(x)|=\ord (h^{\frac23})\left( \int_{-\infty}^0\widetilde U_2(x,t)|v(t)|dt + U_2(x,0)|v(0)|\right),
\end{aligned}
\ee
and therefore,
\be
\label{normK2L}
\pal hK_{2,L}W^* \pal=\ord (h^{\frac23})\sup_{x\in I_L} \int_{-\infty}^0\widetilde U_2(x,t)dt
 + \ord (h^{\frac23})\sup_{x\in I_L} U_2(x,0).
\ee
Moreover, using  the asymptotics of $u_{2,L}^\pm$ and $\partial_xu_{2,L}^\pm$ on $I_L$, 
we see that $ U_2(x,t) =\ord(1)$ uniformly, and fixing some constant $C_1>0$ sufficiently large, we also have,
$$
\widetilde U_2(x,t)  = \left\{
\begin{aligned}
&\ord (h^{-\frac23})\frac{|V_2(t)-E|^{\frac14}e^{-\left| \re \int_t^x  (V_2-E)^{1/2}\right| /h}}{|V_2(x)-E|^{\frac14}}
 &&\ (x,t\leq -C_1h^{\frac23}),\\
&\ord (h^{-\frac 56})|V_2(t)-E|^{\frac14}e^{-\left| \re \int_t^x  (V_2-E)^{1/2}\right| /h}
 &&\ (t\leq -C_1h^{\frac23}\leq x\leq 0),\\
&\ord (h^{-\frac12})|V_2(x)-E|^{-\frac14}e^{-\left| \re \int_0^x  (V_2-E)^{1/2}\right| /h}
 &&\ (x\leq -C_1h^{\frac23}\leq t\leq 0),\\
&\ord (h^{-\frac23})
 &&\ (x,t\in [-C_1h^{\frac23}, 0]).
\end{aligned}
\right.
$$

In particular  $\widetilde U_2(x,t)=\ord(h^{-2/3})$ uniformly, and when $x\leq -\delta$ with $\delta >0$ constant, there exists a positive constant $\alpha$ such that,
$$
\int_{-\infty}^0\widetilde U_2(x,t)dt=\ord(h^{-\frac23})\int_{-\infty}^{-\delta /2}e^{-\alpha |x-t|/h}dt
 +\ord (e^{-\alpha /h})=\ord (h^{\frac13}).
$$
On the other hand, if $\delta$ is chosen sufficiently small and $x\in[-\delta, -C_1h^{2/3}]$, then, there exists a (different) positive constant $\alpha$ such that,
$$
\begin{aligned}
\int_{-\infty}^0\widetilde U_2(x,t)dt &=\int_{-2\delta}^{-C_1h^{2/3}}\widetilde U_2(x,t)dt +\ord (1)\\
&=\ord (h^{-\frac23}|x|^{-\frac14})\int^{2\delta}_{C_1h^{2/3}}t^{\frac14}e^{-\alpha \left| t^{\frac32}-|x|^{\frac32} \right|/h}dt+\ord (1).
\end{aligned}
$$
Setting $t=(hs)^{2/3}$ in the integral, we obtain,
$$
\int_{-\infty}^0\widetilde U_2(x,t)dt=\ord(h^{-\frac23}|x|^{-\frac14}h^{\frac23 + \frac16})\int_1^\infty \frac{e^{-\alpha \left |s-|x|^{\frac 32}/h\right|}}{\sqrt s}ds+\ord(1)=\ord(1).
$$
Finally, when $x\in [-C_1h^{2/3},0]$, we have,
$$
\begin{aligned}
\int_{-\infty}^0\widetilde U_2(x,t)dt &=\int_{-\delta}^{-C_1h^{2/3}}\widetilde U_2(x,t)dt +\ord (1)\\
&=\ord (h^{-\frac56})\int^{\delta}_{C_1h^{2/3}}t^{\frac14}e^{-\alpha t^{\frac32}/h}dt+\ord (1) =\ord(1).
\end{aligned}
$$
Thus, we have proven,
\be
\label{intU2}
\sup_{x\leq 0}\int_{-\infty}^0 \widetilde U_2(x,t)dt =\ord (1),
\ee
and, by (\ref{normK2L}) (and the fact that $U_2=\ord (1)$), (\ref{est3.1.1}) follows.

Now, let us prove the estimate on $M_L:= h^2K_{1,L}WK_{2,L}W^*$. We see on the definition of $K_{1,L}$ and on (\ref{estK2L}) that we have,
\be
\label{decompML}
\begin{aligned}
|M_Lv(x)|= & \ord(h^{\frac43})\int_{-\infty}^0\int_{-\infty}^0 \widetilde U_1(x,t)\widetilde U_2(t,s)|v(s)|ds dt \\
& +\ord(h^{\frac43})\int_{-\infty}^0 \widetilde U_1(x,t)U_2(t,0)|v(0)|dt\\
& +\ord(h^{\frac43})U_1(x,0)\int_{-\infty}^0 \widetilde U_2(0,t)|v(t)|dt\\
& +\ord(h^{\frac43})U_1(x,0)U_2(0,0)|v(0)|.
\end{aligned}
\ee
Using (\ref{intU2}) and the fact that  $U_j=\ord (1)$ uniformly ($j=1,2$), 
we see that the last three terms are $\ord (h^{4/3}) \sup_{I_L}|v|$.

In order to estimate the first term, we use the following properties of $\widetilde U_1$:
For any $\delta>0$ small enough, there exists $\alpha >0$ constant, such that,
$$
\widetilde U_1(x,t) = \left\{
\begin{aligned}
&\ord (h^{-\frac23})e^{-\alpha |t-x|/h} 
&& (x,\ t\leq x^*-\delta)\\
&\ord (e^{-\alpha /h})
&& (t\leq x^*-2\delta,\ x\in[x^*-\delta, 0]),\\
&\ord (e^{-\alpha /h})
&& (x\leq x^*-2\delta,\ t\in[x^*-\delta, 0]),\\ 
&\ord (h^{-\frac56}|t|^{\frac14}) 
&& (x\in [x^*-4\delta,0],\ t\in [-\delta, -C_1h^{\frac23}]),\\ 
&\ord (h^{-\frac23}) 
&& (x\in [x^*-4\delta,0],\ t\in [-C_1h^{\frac23},0]\cup [x^*-4\delta, -\delta]),\\ 
&\ord (h^{-\frac12}|x|^{-\frac14})
&& (t\in [x^*-4\delta,0],\ x\in [-\delta, -C_1h^{\frac23}]),\\ 
&\ord (h^{-\frac23})
&& (t\in [x^*-4\delta,0],\ x\in [-C_1h^{\frac23},0]\cup [x^*-4\delta, -\delta]).
\end{aligned}
\right.
$$

In particular  $\widetilde U_1(x,t)=\ord(h^{-2/3})$ uniformly. 
Moreover, by the properties of $\widetilde U_2$, 
we also know that any part of the integral corresponding to $|t-s|\geq \delta$
 with $\delta >0$ constant is exponentially small.

We first consider the case $x\in (-\infty, x^*-2\delta]$ for a small positive constant $\delta$. 

Then, we see that there exists a constant $\alpha >0$ such that,
\begin{align*}
\iint_{-\infty}^0 \widetilde U_1(x,t)\widetilde U_2(t,s)dtds
 &=\ord (h^{-\frac43})\int_{-\infty}^{x^*-\delta}dt \int_{-\infty}^{x^*-\delta/2} e^{-\alpha (|t-x|+|s-t|)/h}ds\\
&\hskip 6cm+\ord(e^{-\alpha /h})\\
& =\ord(h^{2-\frac43})=\ord(h^{\frac23}).
\end{align*}

Now, when $x\in[x^*-2\delta,0]$, and still denoting by $\alpha$ every new positive constant that may appear, we have,
$$
\iint_{-\infty}^0 \widetilde U_1(x,t)\widetilde U_2(t,s)dtds 
=\int_{x^*-3\delta}^0dt \int_{x^*-4\delta}^0\widetilde U_1(x,t)\widetilde U_2(t,s)ds+\ord(e^{-\alpha /h}),
$$
and,
$$
\int_{x^*-3\delta}^0dt \int_{x^*-4\delta}^0 \widetilde U_1(x,t)\widetilde U_2(t,s)ds 
=\ord(h^{-\frac23})\int_{x^*-3\delta}^{0}dt\int_{x^*-4\delta}^{0}\widetilde U_2(t,s)ds
$$
$$
\begin{aligned}
=\ &\ord(h^{-\frac43})\int_{x^*-3\delta}^{-\delta}dt\int_{x^*-4\delta}^{-\delta/2}e^{-\alpha|t-s|/h}ds +\ord(e^{-\alpha/h})\\
&+\ord(h^{-\frac43})\int_{-\delta}^{-C_1h^{2/3}}dt\int_{-2\delta}^{-C_1h^{2/3}}\frac{|s|^{\frac14}e^{-\alpha \left| |t|^{\frac32}-|s|^{\frac32}\right|/h}}{|t|^{\frac12}}ds\\
&+\ord(h^{-\frac76})\int_{-\delta}^{-C_1h^{2/3}}dt\int_{-C_1h^{2/3}}^0\frac{e^{-\alpha |t|^{\frac32}/h}}{|t|^{\frac14}}ds\\
&+\ord(h^{-\frac32})\int_{-C_1h^{2/3}}^0 dt\int_{-\delta}^{-C_1h^{2/3}}|s|^{\frac14}e^{-\alpha |s|^{\frac32}/h}ds
+\ord(1).
\end{aligned}
$$
Hence,
\be
\label{estinterm1}
\begin{aligned}
& \int_{x^*-3\delta}^0  dt \int_{x^*-4\delta}^0 \widetilde U_1(x,t)\widetilde U_2(t,s)ds\\
 &=\ord(h^{-\frac43})\int^{\delta}_{C_1 h^{2/3}}dt\int^{2\delta}_{C_1 h^{2/3}}\frac{s^{\frac14}e^{-\alpha \left| t^{\frac32}-s^{\frac32}\right|/h}}{t^{\frac12}}ds
+ \ord(h^{-\frac12})\int^{\delta}_{C_1h^{2/3}}\frac{e^{-\alpha t^{\frac32}/h}}{t^{\frac14}}dt\\
&\hskip 2cm  +\ord(h^{-\frac56})\int^{\delta}_{C_1h^{2/3}}s^{\frac14}e^{-\alpha s^{\frac32}/h}ds
+\ord(h^{-\frac13}).
\end{aligned}
\ee
For the first term, the change of variables $(t,s)\mapsto (t^{2/3},s^{2/3})$ gives,
$$
\int^{\delta}_{C_1h^{2/3}}dt\int^{2\delta}_{C_1h^{2/3}}\frac{s^{\frac14}e^{-\alpha \left| t^{\frac32}-s^{\frac32}\right|/h}}{t^{\frac12}}ds
={\mathcal O}(1)\iint_{C_2 h}^{\delta'}\frac{e^{-\alpha |t-s|/h}}{t^{\frac23} s^{\frac16}}dtds,
$$
with $C_2:=C_1^{2/3}$ and $\delta':=(2\delta)^{2/3}$.
Dividing this integral in two parts, depending whether $t\leq s$ or $s\leq t$, and first integrating with respect to the larger of the two variables, we obtain,
\be
\begin{aligned}
\label{estintspec}
\int^{\delta}_{C_1h^{2/3}}dt\int^{2\delta}_{C_1h^{2/3}}\frac{s^{\frac14}e^{-\alpha \left| t^{\frac32}-s^{\frac32}\right|/h}}{t^{\frac12}}ds
 &={\mathcal O}(1)\int_{C_2 h}^{\delta'}\!dt\, \frac{e^{\alpha t/h}}{t^{\frac56}}\int_t^{\delta'}e^{-\alpha s/h}ds\\
 & ={\mathcal O}(h). 
 \end{aligned}
\ee
Moreover, a simple change of variable gives,
$$
\int^{\delta}_{C_1h^{2/3}}\frac{e^{-\alpha t^{\frac32}/h}}{t^{\frac14}}dt=\ord (h^{\frac12}),\quad
\int^{\delta}_{C_1h^{2/3}}{s^\frac14}e^{-\alpha s^{\frac32}/h}ds=\ord (h^{\frac56}),
$$
Inserting into (\ref{estinterm1}), we deduce that, for $x\in[x^*-2\delta,0]$, we have,
\be
\label{estdbleintU2}
\iint_{-\infty}^0 \widetilde U_1(x,t)\widetilde U_2(t,s)dtds =\ord (h^{-\frac13}).
\ee
Finally, going back to (\ref{decompML}), we conclude (\ref{est3.1.2}).

The estimates \eqref{est3.1.3}, \eqref{est3.1.4} for $K_{j,R}$ are proved similarly ($x_\infty$ playing the role of $x^*$).
\hfill$\Box$

Set $M_L:= h^2K_{1,L}WK_{2,L}W^*$ and $M_R:=h^2K_{2,R}W^*K_{1,R}W$.
Thanks to Proposition \ref{claim}, we can define the following four vector-valued functions
as Neumann series for small enough $h$;
\be
\label{w1L}
w_{1,L}:=\left(\begin{array}{c} \sum_{j\geq 0}M_L^ju_{1,L}^-\\-hK_{2,L} W^*\sum_{j\geq 0}M_L^ju_{1,L}^-\end{array}\right),
\ee
\be
\label{w2L}
w_{2,L}:=\left(\begin{array}{c} -\sum_{j\geq 0}M_L^j (hK_{1,L} Wu_{2,L}^-)\\ u_{2,L}^-+hK_{2,L} W^*\sum_{j\geq 0}M_L^j(hK_{1,L} Wu_{2,L}^-)\end{array}\right),
\ee

\be
\label{w1R}
w_{1,R}:=\left(\begin{array}{c} u_{1,R}^-+hK_{1,R}W\sum_{j\geq 0}M_R^j(hK_{2,R}W^*u_{1,R}^-)\\
-\sum_{j\geq 0}M_R^j(hK_{2,R} W^*u_{1,R}^-)
\end{array}\right),
\ee
\be
\label{w2R}
w_{2,R}:=\left(\begin{array}{c} 
-hK_{1,R} W\sum_{j\geq 0}M_R^ju_{2,R}^-\\ 
\sum_{j\geq 0}M_R^ju_{2,R}^-
\end{array}\right).
\ee
It is not difficult to see that they are solutions to the system \eq{sch} and that (see \cite{FMW}, Proposition 4.1),
$$
w_{j,L} \in L^2(I_L)\oplus L^2(I_L)\quad ; \quad w_{j,R} \in L^2(I_R^\theta)\oplus L^2(I_R^\theta).
$$

In order to get the leading term of the imaginary part of resonances, it will be necessary to
compute the asymptotics of these solutions 
up to errors of $\ord (h^{5/3})$. 
This means 
to compute, for example for $w_{1,L}$, two terms $u_{1,L}^-+M_Lu_{1,L}^-$ for the first element, and
one term $-hK_{2,L}W^*u_{1,L}^-$ for the second element just as in the elliptic interaction case.

We will compute the Wronskian $\W_0(E,h)$, which is independent of $x$, at the origin.
Substituting $x=0$ to these solutions or their derivatives, we obtain
the following asymptotic formulae just as in \cite{FMW} (only the remainder estimates are different).
\label{calculen0}
For $S=L,R$, we have, uniformly as $h\to 0_+$,
\be
\label{estw1S}
\begin{aligned}
& w_{1,S} (0) =  \left[\begin{array}{c}
u_{1,S}^-(0) + \beta_{1,S} u_{1,S}^+(0)\\
\alpha_{1,S} u_{2,S}^+(0)
\end{array}\right] +\ord(h^{\frac53});\\
& \widetilde\partial w_{1,S}(0) =\left[\begin{array}{c}
\widetilde\partial u_{1,S}^-(0) + \beta_{1,S} \widetilde\partial u_{1,S}^+(0)\\
\alpha_{1,S} \widetilde\partial u_{2,S}^+(0)
\end{array}\right]  +\ord(h^{\frac53}),
\end{aligned}
\ee
\be
\label{estw2S}
\begin{aligned}
& w_{2,S} (0) =  \left[\begin{array}{c}
\alpha_{2,S} u_{1,S}^+(0)\\
u_{2,S}^-(0) + \beta_{2,S} u_{2,S}^+(0)
\end{array}\right] +\ord(h^{\frac53});\\
& \widetilde\partial w_{2,S}(0) =\left[\begin{array}{c}
\alpha_{2,S} \widetilde\partial u_{1,S}^+(0)\\
\widetilde\partial u_{2,S}^-(0) + \beta_{2,S} \widetilde\partial u_{2,S}^+(0)
\end{array}\right]  +\ord(h^{\frac53}).
\end{aligned}
\ee
Here, $\widetilde\partial$ stands for $h^{2/3}\partial$ and $\alpha_{j,S}$ and $\beta_{j,S}$ are complex numbers defined by
\begin{equation}
\label{defconstjS}
\begin{aligned}
& \alpha_{1,L}=\frac{-\int_{-\infty}^0u_{2,L}^-(t)(W^*u_{1,L}^-)(t)dt}{h\W(u_{2,L}^+, u_{2,L}^-)}; \quad
\beta_{1,L}=\frac{\int_{-\infty}^0u_{1,L}^-(t)(Wr_{1,L})(t)dt}{h\W(u_{1,L}^+, u_{1,L}^-)};\\
& \alpha_{2,L}=  \frac{-\int_{-\infty}^0u_{1,L}^-(t)(Wu_{2,L}^-)(t)dt}{h\W(u_{1,L}^+, u_{1,L}^-)}; \quad 
\beta_{2,L}= \frac{\int_{-\infty}^0u_{2,L}^-(t)(W^*r_{2,L})(t)dt}{h\W(u_{2,L}^+, u_{2,L}^-)};\\
& \alpha_{1,R}=  \frac{-\int^{+\infty}_0u_{2,R}^-(t)(W^*u_{1,R}^-)(t)dt}{h\W(u_{2,R}^-, u_{2,R}^+)}; \quad 
\beta_{1,R}=\frac{\int^{+\infty}_0u_{1,R}^-(t)(Wr_{1,R})(t)dt}{h\W(u_{1,R}^-, u_{1,R}^+)};\\
& \alpha_{2,R}=  \frac{-\int^{+\infty}_0u_{1,R}^-(t)(Wu_{2,R}^-)(t)dt}{h\W(u_{1,R}^-, u_{1,R}^+)}; \quad 
\beta_{2,R}=\frac{\int^{+\infty}_0u_{2,R}^-(t)(W^*r_{2,R})(t)dt}{h\W(u_{2,R}^-, u_{2,R}^+)},
\end{aligned}
\end{equation}
where we have set, for $S=L,R$,
$$
r_{1,S}:=hK_{2,S} W^*u_{1,S}^-,\qquad r_{2,S}:=hK_{1,S}Wu_{2,S}^-,
$$
and where, in the case $S=R$, the integrals run over $I_R^\theta$.

Then we can write the Wronskian $\W_0(E,h)$ up to $\ord(h^{5/3})$ in terms of $\alpha_{j,S}$ and $\beta_{j,S}$ as in Section 6 of \cite{FMW}.
\be
\label{W0modh}
\W_0(E) = -\frac{4\sqrt{2}}{\pi^2}ie^{\frac{\pi i}{4}}\left(\cos \frac{{\mathcal A}}{h}\right)(1 + \ord(h^{\frac43})) 
- \frac{4}{\pi^2} \Bigl( \alpha_{1,R}\alpha_{2,L} + \alpha_{1,L}\alpha_{2,R} \Bigr)
$$
$$+ \frac{\sqrt{2}}{\pi^2} e^{\frac{\pi i}{4}}\left(\sin \frac{{\mathcal A}}{h}\right)
\Bigl( 4i \alpha_{1,R}\alpha_{2,R} - \alpha_{1,L}\alpha_{2,L} + 2i (\beta_{1,R} + \beta_{1,L}) \Bigr)
+ \ord (h^{\frac53}),
\ee
where $\mathcal A$ is the action defined in \eqref{action}.
Notice that,  in (6.2) of \cite{FMW},  we used the facts $\alpha_{1,S}=\alpha_{2,S}$ and $\beta_{1,S}=\beta_{2,S}$
at the principal level.

The constants $\alpha_{j,S}$ and $\beta_{j,S}$  have the following estimates:
\begin{proposition}\sl
\label{calculen0suite}
Let $E=\rho h^{2/3} \in {\mathcal D_h}(C_0)$. Then one has, as $h\to 0$,
$$
\alpha_{j,S}=\ord(h^{\frac23}),\quad \beta_{j,S}=\ord(h^{\frac43}),\quad j=1,2,\ S=L,R.
$$
More precisely, one has
$$
\label{approxalphabeta}\begin{aligned}
&\alpha_{1,R} = \frac{e^{\pi i/4}}{\sqrt{2}}\pi r_1(0)h^{\frac23} 
 \Bigl( \nu^A_{1,R}(\textnormal{Re}\,\rho) -i \nu^B_{1,R}(\textnormal{Re}\,\rho) \Bigr) + \ord(h),\\[7pt]
&\alpha_{2,R} = \frac{e^{\pi i/4}}{\sqrt{2}}\pi r_1(0) h^{\frac23} \Bigl( \nu^A_{2,R}(\textnormal{Re}\,\rho)
 -i \nu^B_{2,R}(\textnormal{Re}\,\rho) \Bigr)+ \ord(h),\\
&\alpha_{1,L} = 2\pi r_1(0) h^{\frac23} \left\{\left(\sin \frac{{\mathcal A}}{h}\right) \nu^A_{1,L}(\textnormal{Re}\,\rho)
 + \left(\cos \frac{{\mathcal A}}{h}\right) \nu^B_{1,L}(\textnormal{Re}\,\rho)\right\} + \ord(h), \\[7pt]
&\alpha_{2,L} = 2\pi r_1(0) h^{\frac23} \left\{\left(\sin \frac{{\mathcal A}}{h}\right) \nu^A_{2,L}(\textnormal{Re}\,\rho)
 + \left(\cos \frac{{\mathcal A}}{h}\right) \nu^B_{2,L}(\textnormal{Re}\,\rho)\right\} + \ord(h),\\
&\im\, \beta_{1,R} = \pi^2r_1(0)^2h^{\frac43} \Bigl( \nu^A_{1,R}(\textnormal{Re}\,\rho)
\nu^A_{2,R}(\textnormal{Re}\,\rho)
 + \nu^B_{1,R}(\textnormal{Re}\,\rho)\nu^B_{2,R}(\textnormal{Re}\,\rho) \Bigr) +\ord(h^{\frac53}),\\
& \im\, \beta_{1,L} = \ord(h^{\frac53}),
\end{aligned}
$$
where
$$
\nu^A_{1,R} (t) = \int_0^\infty \ai'(\tau_1^{1/3}(y-\frac t{\tau_1}))\, \ai (-\tau_2^{1/3}(y+\frac t{\tau_2}))dy,
$$
$$
\nu^B_{1,R}  (t) = \int_0^\infty \ai'(\tau_1^{1/3}(y-\frac t{\tau_1})) \,\bi(-\tau_2^{1/3}(y+\frac t{\tau_2}))dy,
$$
$$
\nu^A_{2,R} (t) = \int_0^\infty \ai(\tau_1^{1/3}(y-\frac t{\tau_1}))\, \ai' (-\tau_2^{1/3}(y+\frac t{\tau_2}))dy,
$$
$$
\nu^B_{2,R}  (t) = \int_0^\infty \ai(\tau_1^{1/3}(y-\frac t{\tau_1})) \,\bi' (-\tau_2^{1/3}(y+\frac t{\tau_2})) dy,
$$
$$
\nu^A_{1,L} (t) = \int^0_{-\infty} \ai'(\tau_1^{1/3}(y-\frac t{\tau_1}))\, \ai (-\tau_2^{1/3}(y+\frac t{\tau_2}))dy,
$$
$$
\nu^B_{1,L}  (t) = \int^0_{-\infty} \bi'(\tau_1^{1/3}(y-\frac t{\tau_1})) \,\ai (-\tau_2^{1/3}(y+\frac t{\tau_2})) dy,
$$
$$
\nu^A_{2,L} (t) = \int^0_{-\infty} \ai(\tau_1^{1/3}(y-\frac t{\tau_1}))\, \ai' (-\tau_2^{1/3}(y+\frac t{\tau_2})) dy,
$$
$$
\nu^B_{2,L}  (t) = \int^0_{-\infty} \bi(\tau_1^{1/3}(y-\frac t{\tau_1})) \,\ai' (-\tau_2^{1/3}(y+\frac t{\tau_2})) dy.
$$
\end{proposition}

\noindent{\bf Proof:}\quad 
We only prove the formula for  $\alpha_{1,L}$. Other formulas can be obtained similarly.

Thanks to the exponential decay of $u_{2,L}^-$ away from 0, we obtain,
\begin{align*}
\alpha_{1,L} &= \frac{\pi h^{\frac23}}{2}\int_{-\delta}^0 u_{2,L}^-(t) ( r_1(t) u_{1,L}^-(t) )' dt + \ord(h),
\end{align*}
where $\delta >0$ is arbitrarily small. 
We divide the integral into three parts, introducing a large parameter $\lambda$ satisfying 
$\lambda h^{2/3}\to 0$;
$$
\int_{-\delta}^0 u_{2,L}^-(t) ( r_1(t) u_{1,L}^-(t) )' dt=I_1+I_2+I_3,
$$
where 
$$
I_1 = \int_{-\delta}^{-\lambda h^{2/3}} u_{2,L}^-(t) r_1(t) (u_{1,L}^-)'(t) dt,\quad
I_2 =\int_{-\lambda h^{2/3}}^0 u_{2,L}^-(t) r_1(t) (u_{1,L}^-)'(t) dt, 
$$
$$
I_3 = \int_{-\delta}^0 u_{2,L}^-(t) r_1'(t) u_{1,L}^-(t) dt.
$$
On $[-\delta, -\lambda h^{2/3}]$, both $u_{2,L}^-$ and $(u_{1,L}^-)'$ are
 of the WKB form, one of which is exponentially small and the other of which is oscillating; for some $c>0$,
$$
| u_{2,L}^-(t)|\le \frac{h^{\frac16}}{|t|^{\frac14}}e^{-c|t|^{\frac32}/h},\quad
| (u_{1,L}^-)'(t)|\le c h^{-\frac56}|t|^{\frac14},
$$
and it follows that
$$
|I_1| = \ord(h^{-\frac23}) \int_{-\delta}^{-\lambda h^{2/3}} e^{-c|t|^{\frac32}/h} dt
 = \ord(e^{-c\lambda^{\frac32}}).
$$
Taking $\lambda$ larger than $((3c)^{-1}|\ln h|)^{2/3}$, we get $I_1=\ord(h^{1/3})$.

On the other hand, using Proposition 5.1, A.2 and A.5 of \cite{FMW} about the asymptotic properties of the solutions
$u_{2,L}^-$ and $u_{1,L}^-$ near the crossing point, we have,
$$
\begin{aligned}
I_2 &= 4h^{-\frac23} \int_{-\lambda h^{2/3}}^0 
r_1(t)\left(\frac{\xi_1'}{\xi_2'}\right)^{-\frac12} \ai (-h^{-\frac23}\xi_2) \times \\
& \qquad \left(
(\sin\frac{\mathcal A}{h})\ai' (h^{-\frac23}\xi_1) + (\cos\frac{\mathcal A}{h})\bi' (h^{-\frac23}\xi_1) 
\right) dt +\ord(h^{\frac13}).
\end{aligned}
$$
Making the change of variable $y:=h^{-2/3}t$ and $\rho := h^{-2/3}E$ as in \cite{FMW},  we obtain,
$$
\begin{aligned}
  I_2 &= 4 r_1(0) \int_{-\lambda}^0 \ai(-y-\rho)
\left\{\left(\sin \frac{{\mathcal A}}{h}\right) 
\ai'(y-\rho) + \left(\cos \frac{{\mathcal A}}{h}\right) \bi'(y-\rho)\right\} dy \\
 &\quad + \ord(h^{\frac23}\lambda^2) + \ord(h^{\frac13}).
\end{aligned}
$$
Then, using the behaviour of Airy functions at $\pm\infty$, this leads to,
$$
\begin{aligned}
I_2
 &= 4r_1(0) \int_{-\infty}^0 \ai(-y-\rho)
\left\{\left(\sin \frac{{\mathcal A}}{h}\right) \ai'(y-\rho) + \left(\cos \frac{{\mathcal A}}{h}\right) \bi'(y-\rho)\right\} dy \\
&\quad  +\ord(e^{-c'\lambda^{\frac32}}) + \ord(h^{\frac23}\lambda^2) + \ord(h^{\frac13}),
\end{aligned}
$$
with $c'>0$ constant. Hence, taking $\lambda:=((3c'')^{-1}|\ln h |)^{2/3}$ with $c''=\min \{c,c'\}$, 
the error is $\ord (h^{1/3})$.

The integral $I_3$ contains $u_{1,L}^-$ instead of its derivative compared with $I_1$ and $I_2$. 
Then one easily see that $I_3=\ord(h^{2/3})$.

Thus the formula for $\alpha_{1,L}$ is obtained.
\hfill$\Box$

This proposition \ref{calculen0suite} together with  \eqref{W0modh} imply that 
there exists a bounded complex-valued function $G(E,h)$ of 
$E=\rho h^{2/3} \in {\mathcal D_h}(C_0)$ and $h$ sufficiently small such that 
$E \in {\mathcal D_h}(C_0)$ is a resonance of $P$ if and only if,
\begin{equation}
\label{BS}
\cos \frac{{\mathcal A}(E)}{h}  
= h^{\frac43}\left (\sin \frac{{\mathcal A}(E)}{h}\right )G(E,h).
\end{equation}
More precisely, 
from the fact $\sin^2 ({\mathcal A}/h) = 1+\ord (h^{8/3})$,
we obtain the following asymptotic formula for the imaginary part of $G(E,h)$: As $h\to 0$,
$$
\im\, G =  \pi^2 r_1(0)^2\left( \nu^A_{1,R} (\re \rho)+ \nu^A_{1,L}(\re \rho) \right)\left( \nu^A_{2,R} (\re \rho)
+ \nu^A_{2,L}(\re \rho) \right) + \ord(h^{\frac13}).
$$

Finally notice that the functions
\begin{equation}\label{airyint}
\begin{aligned}
\nu^A_{1,R}(t) + \nu^A_{1,L}(t)
 &= \int_{-\infty}^\infty \ai'(\tau_1^{\frac{1}{3}} (y - \frac t{\tau_1 })) \ai(-\tau_2^{\frac{1}{3}} (y + \frac t{\tau_2}))\,dy,\\
\nu^A_{2,R}(t) + \nu^A_{2,L}(t)
 &= \int_{-\infty}^\infty \ai(\tau_1^{\frac{1}{3}} (y - \frac t{\tau_1 })) \ai'(-\tau_2^{\frac{1}{3}} (y + \frac t{\tau_2})) \,dy,
\end{aligned}
\end{equation}
are both proportional to the derivative of the function
$$
\int_{-\infty}^\infty \ai(\tau_1^{\frac{1}{3}} (y - \frac t{\tau_1 })) \ai(-\tau_2^{\frac{1}{3}} (y + \frac t{\tau_2}))\,dy
 = (\tau_1 + \tau_2)^{-\frac{1}{3}} \ai ( -\tau_3^{-\frac{2}{3}}t ), 
$$
with $\tau_3^{-1} = \tau_1^{-1} + \tau_2^{-1}$, and their product is given by
$ \frac{\tau_3^{\frac{1}{3}}}{\tau_1 + \tau_2} 
\left(\ai' (- \tau_3^{-\frac{2}{3}}\lambda_k )\right)^2$.

Theorem \ref{mainth2} then follows from \eqref{BS}.


\end{document}